\documentclass[aps,twocolumn,floatfix,prl,superscriptaddress,10pt]{revtex4-1}
\usepackage{graphicx,amsmath,bbm}

\newcommand{\ua}{\uparrow}
\newcommand{\da}{\downarrow}

\def\braket#1{\mathinner{\langle{#1}\rangle}}
\def\bra#1{\left\langle#1\right|}
\def\ket#1{\left|#1\right\rangle}

\begin{document}

\title{Nuclear spin pumping and electron spin susceptibilities}

\author{J.\ Danon}
\affiliation{{Dahlem Center for Complex Quantum Systems, Freie Universit\"{a}t Berlin, Arnimallee 14, 14195 Berlin, Germany}}
\author{Yu.\ V.\ Nazarov}
\affiliation{Kavli Institute of Nanoscience, Delft University of Technology, 2600 GA Delft, The Netherlands}

\begin{abstract}
In this work we present a
new 
formalism to evaluate the nuclear spin dynamics driven by hyperfine interaction with non-equilibrium electron spins. To describe the dynamics up to second order in the hyperfine coupling, it suffices to evaluate the susceptibility and fluctuations of the electron spin. Our approach does not rely on a separation of electronic energy scales or the specific choice of electronic basis states, thereby overcoming practical problems which may arise
in certain limits
when using a more traditional formalism based on rate equations.
\end{abstract}

\maketitle

\section{Introduction}

In recent years, considerable theoretical and experimental work is aimed at the controlled manipulation of electron spins in nanoscale solid state devices~\cite{spinrev}. This research is motivated by actual applications, such as in digital information storage and read-out~\cite{wolfscience,springerlink:10.1007}, but also by the possibility of using the spin of electrons as computational units (qubits) in a quantum computer~\cite{qdqb}.

One of the mechanisms influencing the electron spin dynamics in these nano-devices is the hyperfine interaction between the electron spin and the nuclear spins of the device's constituent material. For spin qubits hosted in semiconductor quantum dots~\cite{ronaldrev}, hyperfine interaction has been identified as the main source of decoherence, causing the spin coherence time to be in the ns range~\cite{frank:science,petta:science,klg}. Much recent experimental and theoretical work is aimed at suppressing this hyperfine induced decoherence~\cite{reillyt2,PhysRevLett.102.057601}. In other semiconductor nano-structures~\cite{onosb,greilich:science} --- possibly also metallic structures~\cite{PhysRevLett.97.146602} --- hyperfine interaction can even dominate the electronic transport properties and spin dynamics. Understanding the role of hyperfine interaction in spintronic devices therefore is crucial in the development of the field.

Traditionally, the interplay between electronic and nuclear spins is treated in the convenient framework of rate equations~\cite{opticalorientation,dyakonovbook}. The rates of hyperfine transitions flipping nuclear spins `up' and `down' are separately calculated using Fermi's golden rule, and the balancing of these rate yields the net nuclear spin pumping in the system. Although this approach works very well in many cases~\cite{tartakovskii:prl,Rudner}, it can become cumbersome when the electron spin dynamics are more complicated. Strong dynamical nuclear spin effects have been observed in GaAs double quantum dots in the spin blockade regime under conditions of electron spin resonance~\cite{ivo:nature}. Nuclear spin flips in this setup are due to second order processes, and some of the transitions have a vanishingly small energy difference between initial and virtual state, which cannot be dealt with in a standard Fermi golden rule approach~\cite{PhysRevLett.103.046601}. The situation is even worse for systems with strong spin-orbit coupling. In InAs nanowire double quantum dots in the spin blockade regime, signatures of strong dynamic polarization have been observed too~\cite{pfund:prl}. In this case, the presence of too many comparable energy scales makes it impossible to choose a unique set of basis states to describe the electron dynamics in~\cite{PhysRevB.80.041301}. It is clear that a basis-independent description of the coupled electron-nuclear spin dynamics, not involving any separate transition rates, is highly desirable.

In this work we present an alternative method to evaluate the spin dynamics of electrons and nuclei coupled to each other by hyperfine interaction. We show that, in order to describe the nuclear spin dynamics up to second order in the hyperfine coupling, it suffices to evaluate the fluctuations and susceptibility of the electron spins in the system, which can be done using linear response theory. Our approach does not rely on the calculation of separate transition rates with Fermi's golden rule, nor on the choice of electronic basis states to work in.
As a result, our formalism may be 
applicable in cases where Fermi's golden rule cannot be used, and thus provides a useful alternative.

Since a Fermi's golden rule (FGR) calculation often has to be adapted to specific circumstances~\cite{opticalorientation,dyakonovbook,tartakovskii:prl,Rudner}, it is difficult to pick a single implementation of it and make a good comparison to our approach. Therefore we will illustrate our ideas with two example systems: (i) We will show how in a simplest toy system our formalism and a smart implementation of FGR produce identical results. (ii) We will investigate a more complicated system, which cannot be dealt with in an FGR approach, and show how our formalism straightforwardly produces an equation for the hyperfine driven nuclear spin dynamics.

\section{Main result}
Let us start with presenting the main result of our work: an equation for the hyperfine driven dynamics of the expectation value of a nuclear spin coupled to (many) electronic spins. Under conditions which we will specify below, this equation takes the closed form
\begin{equation}
\frac{d\langle \hat K^a\rangle}{dt} = \varepsilon^{abc}S^b\langle\hat K^c\rangle- \frac{1}{2}Q^{ab} \langle \hat K^b\rangle+ \frac{1}{3} P^a ,
\label{eq:dkdtfin}
\end{equation}
where $\{a,b,c\} \in \{x,y,z\}$, and we use the convention that over repeated indices still has to be summed.
The operator $\hat{\bf K}$ is the nuclear spin operator for the nucleus under consideration, and $\varepsilon^{abc}$ is the permutation tensor. The vectors ${\bf S}$ and ${\bf P}$, and the matrix $Q$ are defined as
\begin{equation}
\begin{split}
S^a & = Av_0 \langle \hat S^a ({\bf r}_n,t)\rangle\\
Q^{ab} & = (Av_0)^2 (\delta^{ab} R^{cc} - R^{ba})\\
P^a & = (Av_0)^2 K(K+1) \varepsilon^{abc} \chi^{bc} ,
\end{split}\label{eq:rchi}
\end{equation}
$A$ being the material-specific hyperfine coupling energy, $1/v_0$ the density of nuclei, $K$ the total nuclear spin, ${\bf r}_n$ the position of the nucleus, and $\hat{{\bf S}}$ the electron spin density operator, $\hat{{\bf S}}({\bf r},t) = \tfrac{1}{2}\hat \psi^\dagger_\alpha ({\bf r},t) {\boldsymbol \sigma}_{\alpha\beta} \hat \psi_\beta ({\bf r},t)$. Note that we have set for transparency $\hbar=1$, or, in other words, we express all energies in terms of corresponding frequencies.
We further assumed for simplicity only one type of spin-carrying nuclei present, having a constant density.
The symbols $\chi$ and $R$ represent correlation functions of the local electronic spin density,
\begin{equation}
\begin{split}
R^{ab} & = \phantom{-i}\int^t \langle [\hat S^a({\bf r}_n,t),\hat S^b({\bf r}_n,t')]_+\rangle_c dt'\\
\chi^{ab} & = -i \int^t \langle [ \hat S^a({\bf r}_n,t),\hat S^b({\bf r}_n,t')]_-\rangle_c dt',
\end{split}
\label{eq:chir}
\end{equation}
where the square brackets denote the (anti)commutator of two operators, i.e., $[\hat A,\hat B]_\pm = \hat A\hat B\pm \hat B\hat A$, and the subscript $c$ means that we have to use the `connected' part of the expectation value, i.e., remove the contribution of the averages $\langle AB \rangle_c = \langle AB \rangle - \langle A \rangle\langle B\rangle$.

Note that, since hyperfine interaction works two-way, the electronic dynamics, and thereby ${\bf S}$, $Q$ and ${\bf P}$, can in turn also be affected by the state of the nuclear spins in the system. This creates a feedback mechanism in Eq.\ (\ref{eq:dkdtfin}), which makes the equation non-linear. However, in all practical implementations, one can make use of the large difference in time scales of the electronic and nuclear spin dynamics. This allows to treat the ensemble of nuclear spins effectively as a static `classical' magnetic field when evaluating the electronic correlators.

The first term in Eq.\ (\ref{eq:dkdtfin}) is first order in the hyperfine coupling $A$ and gives rise to a precession of the nuclear spin around the direction of the time-averaged local electron spin polarization. The precession frequency depends on the magnitude of this polarization $|v_0 \langle {\hat{\bf S}}({\bf r}_n,t)\rangle|$, as well as on the strength $A$ of the hyperfine coupling.

The other two terms are second order in $A$, and describe the effect on the nucleus of fluctuations of electron spin. The correlation functions $R^{ab}$ in the matrix $Q$ are ordered as the `classical' noise of electron spin ${\hat{\bf S}}({\bf r}_n,t)$. It persists even if the quantum operator of spin is replaced with a classical fluctuating field affecting the nuclear spin. Since such fluctuations cause random rotation of this spin and therefore isotropization of its density matrix, the effect of $R^{ab}$ is relaxation of the nuclear spin. The correlators $\chi^{ab}$ in the vector ${\bf P}$ have the same structure as a Kubo formula and give the stationary susceptibility, that is, the response of $\hat S_a$ in the point ${\bf r}_n$ to a constant magnetic field in the $b$-direction, concentrated in the same point. One can interpret the $\chi^{ab}$ as the `quantum' noise of electron spin: the part of the fluctuations in ${\hat{\bf S}}({\bf r}_n,t)$ which solely exists due to the non-commutativity of the spin operators. It is worth to note that pumping of the nuclear spin is produced by the asymmetric part of the tensor $\chi^{ab}$ only. The Onsager relations state that susceptibilities are symmetric for an equilibrium system, $\chi^{ab}=\chi^{ba}$. Therefore, pumping can only take place if the electron part of the system is driven out of equilibrium by some external agent.

In order to find an explicit expression for $d\langle \hat{\bf K}\rangle/dt$ as given by Eq.\ (\ref{eq:dkdtfin}), one needs to find the elements of ${\bf S}$, $Q$ and ${\bf P}$. The first step is to find the steady-state electron spin density $\langle \hat{\bf S}({\bf r}_n) \rangle$, which immediately yields the vector ${\bf S}$.
The elements of $Q$ and ${\bf P}$ correspond to the local fluctuations and susceptibility of this electron spin density, and can be evaluated using linear response theory, as we will illustrate below. Let us emphasize here that none of the steps in such a calculation involves the evaluation of any separate transition rate between different electronic-nuclear levels or depends on the specific choice of electronic basis states.

\section{Derivation}

We will now show how to derive Eq.\ (\ref{eq:dkdtfin}) from a second order perturbation theory in the hyperfine interaction.
We start from an effective equation of motion for the reduced density matrix $\hat \rho^k$ of the nuclear spins which reads up to second order in the perturbation $\hat H_\text{hf}(t)$ (see the Appendix for a derivation)
\begin{equation}
\begin{split}
\frac{d\hat \rho^k}{dt} = \overline{\mathrm{Tr}}_e\Big\{& - i [\hat H_\text{hf}(t),\hat \rho^e_0\otimes\hat\rho^k ]_- \\
& - \int^t [ \hat H_\text{hf}(t), [ \hat H_\text{hf}(t'),\hat \rho^e_0\otimes\hat\rho^k ]_-]_- dt'\Big\},
\end{split}
\label{eq:2ohf}
\end{equation}
where $\hat\rho^e_0$ is the electronic part of the (quasi-)stationary density matrix described by the unperturbed Hamiltonian $\hat H_0$. This Hamiltonian includes all processes and degrees of freedom which are relevant for the electron spin dynamics, such as (e.g.\ in a transport setup) all leads and tunnel couplings to these leads. The trace $\overline{\mathrm{Tr}}_e\{\dots\}$ denotes a trace over the electronic degrees of freedom, where the horizontal bar indicates that from the resulting correlation functions only the fully connected contribution should be taken into account (see Appendix).

The perturbation $\hat H_\text{hf}(t)$ accounts for the hyperfine interaction which couples the nuclear spins to the local electron spin density. Assuming that the electrons of interest have $s$-type orbitals, we use
\begin{equation}
\hat H_\text{hf}(t) = Av_0 \sum_n \hat S^a({\bf r}_n,t)\hat K^a_n(t),
\label{eq:hfint}
\end{equation}
where the index $a$ determines, as above, the spin component of the operators, $a \in \{x,y,z\}$. The operators are represented in the interaction picture (see the Appendix), which means that their time-dependence is governed by $\hat H_0$, i.e., $\hat{{\bf S}}({\bf r}_n,t) = e^{i\hat H_0 t}\hat{{\bf S}}({\bf r}_n)e^{-i\hat H_0t}$. We assume that the nuclear spins evolve on a time scale which is much longer than that of the electron spin dynamics, so that for all times of interest we can use $\hat{{\bf K}}_n(t) = \hat{{\bf K}}_n$. We see that, if we focus on the dynamics of one specific nuclear spin (as we will do below), our perturbative approach requires that $Av_0$ is much smaller than the energy scales in the unperturbed Hamiltonian $\hat H_0$.

We then substitute the perturbation (\ref{eq:hfint}) into the equation of motion (\ref{eq:2ohf}), make use of the fact that a trace is invariant under cyclic permutation of the operators in its argument, and then finally arrive at
\begin{widetext}
\begin{equation}
\begin{split}
\frac{d\hat \rho^k}{dt} = & -i Av_0 \sum_n \langle \hat S^a({\bf r}_n,t)\rangle [\hat K^a_n,\hat \rho^k]_- \\
& -(Av_0)^2 \sum_{n,m} \bigg\{ \int^t \langle \hat S^a({\bf r}_n,t)\hat S^b({\bf r}_m,t')\rangle_c dt' [\hat K^a_n ,\hat K^b_m \hat\rho^k ]_- - \int^t \langle \hat S^b({\bf r}_m,t')\hat S^a({\bf r}_n,t)\rangle_c dt' [\hat K^a_n,\hat\rho^k\hat K^b_m ]_-\bigg\},
\end{split}
\label{eq:drkdt2}
\end{equation}
\end{widetext}
where now brackets are used to denote the trace over the electronic part of the density matrix, $\langle \dots\rangle \equiv \mathrm{Tr}_e \{ \dots \hat\rho^e_0 \}$ in general, and $\langle \dots\rangle_c \equiv \overline{\mathrm{Tr}}_e \{ \dots \hat\rho^e_0 \}$ for the fully connected terms. From (\ref{eq:drkdt2}) we can derive equations of motion for the expectation values of the nuclear spin operators.

\subsection{A single nucleus, $K=1/2$}
Let us first focus on the transparent case of one single nucleus with nuclear spin $K = 1/2$, interacting with a local electron spin density.
Up to first order, i.e., using only the first term in (\ref{eq:drkdt2}), we find
\begin{equation}
\begin{split}
\frac{d \langle {\hat{K}^a}\rangle}{dt}^{\!(1)} \!\!\! = \mathrm{Tr}\bigg\{\hat{{K}}^a \frac{d\hat\rho^k}{dt}\bigg\} & = -i S^b \mathrm{Tr}\big\{[\hat K^a,\hat K^b]_-\hat\rho^k \big\}\\ &  = \varepsilon^{abc} {S}^b\langle \hat{{K}}^c\rangle,
\end{split}
\label{eq:dkdt1}
\end{equation}
describing a precession of the nuclear spin around the local electron spin density at the position ${\bf r}_n$ of the nucleus, with the vector ${\bf S}$ defined as ${\bf S}\equiv Av_0\langle \hat{\bf S}({\bf r}_n,t)\rangle$. Obviously, the traces appearing in (\ref{eq:dkdt1}) and the corresponding averages $\langle \dots \rangle$ involve tracing and averaging over the nuclear spin part of the system only. Since no confusion is possible, we do not label these traces and averages separately with a $k$.

The exchange of angular momentum between electrons and the nucleus is to leading order described by the second term in (\ref{eq:drkdt2}), i.e., second order in $\hat H_\text{hf}$. We write the correlation functions of the electron spin in terms of the quantities $R^{ab}$ and $\chi^{ab}$ introduced in (\ref{eq:chir}), and make again use of the cyclic invariance of the trace. For the convenient case of $K=1/2$, we can simplify products of nuclear spin operators using $\hat K^a\hat K^b = \tfrac{i}{2}\varepsilon^{abc}\hat K^c + \tfrac{1}{4}\delta^{ab}$. Along these lines we simplify the second order term,
\begin{widetext}
\begin{equation}
\begin{split}
\frac{d \langle {\hat{K}}^a\rangle}{dt}^{\!(2)} \!\!\! & =-\frac{(Av_0)^2}{2}\Big\{ (R^{bc}+i\chi^{bc} )
\mathrm{Tr}\big\{ \hat K^a\hat K^b\hat K^c\hat\rho^k - \hat K^a\hat K^c\hat\rho^k\hat K^b\big\} - (R^{bc}-i\chi^{bc} )
\mathrm{Tr}\big\{ \hat K^a\hat K^b\hat\rho^k\hat K^c - \hat K^a\hat\rho^k\hat K^c\hat K^b\big\}\Big\} \\
& = -\frac{(Av_0)^2}{2}\Big\{ R^{bc}\big\langle [[\hat K^a,\hat K^b]_-,\hat K^c]_-\big\rangle  + i\chi^{bc} \big\langle [[\hat K^a,\hat K^b]_-,\hat K^c]_+\big\rangle \Big\}\\
& = \phantom{-}\frac{(Av_0)^2}{2}\Big\{R^{ba}\langle {\hat{K}}^b\rangle-R^{bb}\langle\hat{K}^a\rangle+ \varepsilon^{abc} \tfrac{1}{2}\chi^{bc}\Big\},
\end{split}
\label{eq:dkdt2}
\end{equation}
which is, combined with Eq.\ (\ref{eq:dkdt1}), identical to the expression given in Eq.\ (\ref{eq:dkdtfin}).

\subsection{Many nuclei, $K>1/2$}

When following the same derivation for the case of many nuclei and spin higher than $K = 1/2$, it is generally not possible to derive a closed set of equations for $d\langle \hat{\bf K}_n\rangle/dt$. The first order equations do not change and are still given by Eq.\ (\ref{eq:dkdt1}). The second order equation however does change, and becomes
\begin{equation}
\frac{d \langle {\hat{K}}^a_n\rangle}{dt}^{\!(2)} \!\!\! = \, 
\frac{(Av_0)^2}{2}\Big\{  R^{ba}\langle {\hat{K}}^b_n\rangle-R^{bb}\langle\hat{K}^a_n\rangle
 + \varepsilon^{abc} \chi_{nn}^{bd}\langle [\hat{K}^d_n, \hat{K}^c_n]_+\rangle +2 \sum_{m\neq n}\varepsilon^{abc} \chi_{nm}^{bd}\langle \hat{K}^d_m \hat{K}^c_n \rangle \Big\},
\label{eq:dkdt3}
\end{equation}
\end{widetext}
where the correlator $\chi$ has acquired two more indices,
\begin{equation}
\chi^{ab}_{nm} = -i \int^t \langle [ \hat S^a({\bf r}_n,t),\hat S^b({\bf r}_m,t')]_-\rangle_c dt'.
\label{eq:chi2}
\end{equation}
This correlation function now describes the {\it non-local} susceptibility of the electron spin density, i.e., the linear response in the spin coordinate $\hat S^a$ at the position ${\bf r}_n$ due to a perturbation along $\hat S^b$ at another position ${\bf r}_m$.

We notice two practical problems with Eq.~(\ref{eq:dkdt3}). First of all, for $K > 1/2$ we cannot use the relation $\hat K_n^a\hat K_n^b = \tfrac{i}{2}\varepsilon^{abc}\hat K_n^c + \tfrac{1}{4}\delta^{ab}$ to simplify products of nuclear spin operators, and secondly, also correlations $\langle \hat{K}^d_m \hat{K}^c_n \rangle$ between {\it different} nuclei play a role. This makes it impossible to derive from Eq.~(\ref{eq:dkdt3}) a closed set of equations for $d\langle \hat{\bf K}_n\rangle/dt$.

However, typically the number of nuclear spins in the ensemble is large ($N\sim 10^4$--$10^6$). When focusing on the dynamics of one single nuclear spin, as we do in Eq.~(\ref{eq:dkdt3}), the presence of the large number of other nuclear spins can, to first approximation, be considered as a static `classical' field. In other words, we can assume the fluctuations in the nuclear field $\langle \hat{K}^d_m \hat{K}^c_n \rangle_c$ to be much smaller than the average values of the field $\langle \hat{K}^d_m\rangle\langle \hat{K}^c_n \rangle$. This assumption then allows us to replace $\langle \hat{K}^d_m \hat{K}^c_n \rangle \to \langle \hat{K}^d_m \rangle\langle\hat{K}^c_n \rangle$ in the last term of Eq.~(\ref{eq:dkdt3}). We emphasize here that in a standard FGR calculation, the same assumption is implicitly used: Transition rates between different spin states of a single nucleus are evaluated, ignoring any possible corrections due to correlations between this spin and the other nuclear spins.

With this simplification, the last term of Eq.~(\ref{eq:dkdt3}) now reads $(Av_0)^2\sum_{m} \varepsilon^{abc} \chi_{nm}^{bd}\langle \hat{K}^d_m \rangle\langle\hat{K}^c_n \rangle$ and describes an effective electron mediated internuclear spin-spin coupling. This effect can be interpreted as a correction to the first-order equation (\ref{eq:dkdt1}): The effective field around which the $n$-th nuclear spin precesses is adjusted $S^a \to S^a  + (Av_0)^2\sum_{m} \chi^{ab}_{nm}\langle\hat{{K}}^b_m \rangle$, i.e., it acquires a contribution due to the effective field produced by all other nuclear spins, mediated by the electron spin~\cite{footnote2}.

The third term in (\ref{eq:dkdt3}) contains products of spin operators of the {\it same} nucleus, and must thus be treated on different footing. In most realistic experimental circumstances, the temperature is much larger than the nuclear Zeeman energy. This means that the (thermal) equilibrium density matrix of the nuclear spins can be approximated to be isotropic, i.e., $\hat\rho^k_0 \propto \mathbbm{1}$. As long as we focus on small deviations from this equilibrium, i.e., small polarizations $\langle\hat K^a\rangle$, the nuclear spin density matrix is always close to isotropic. In this case, quadratic terms like $\langle \hat K_n^a \hat K_n^a\rangle_c$ will dominate the contribution from $\langle [\hat{K}^d_n, \hat{K}^c_n]_+\rangle_c$ in (\ref{eq:dkdt3}), since they are of order unity, $\langle (\hat K_n^a)^2\rangle_c \approx \tfrac{1}{3}K(K+1)$. We thus can concentrate on these quadratic contributions, and approximate $\langle \hat K_n^c \hat K_n^d\rangle \approx \frac{1}{3}\delta^{cd}K(K+1)$. We see that then finally Eq.~(\ref{eq:dkdt3}) reduces to the simple result of Eq.~(\ref{eq:dkdt2}).

Let us repeat here that this reduction is based on two assumptions: (i) we neglect the correlations between the nuclear spin under consideration and all other nuclear spins (as is also done in an FGR calculation), and (ii) we only allow for small deviations of $\hat\rho^k$ from equilibrium, so that $\langle \hat K_n^c \hat K_n^d\rangle \approx \frac{1}{3}\delta^{cd}K(K+1)$. In fact, one does not necessarily have to use these assumptions and make this simplification. Equation (\ref{eq:drkdt2}) gives the time-evolution up to second order in $\hat H_\text{hf}$ of the full nuclear spin density matrix in terms of the electron spin fluctuations and susceptibilities. One could solve this Equation (numerically) for any initial condition $\hat\rho^k_0$ and calculate the time-dependent expectation value of any desired nuclear spin operator.

\section{Implementation}

With the formalism presented in Eqs (\ref{eq:dkdtfin})--(\ref{eq:chir}), we are able to express the hyperfine driven nuclear spin dynamics in terms of the susceptibility and fluctuations of the electron spin density. Let us now explain how one could calculate the necessary electronic correlation functions using linear response theory. For a given setup, we write down the full set of Bloch equations describing the evolution of the electronic density matrix
under $\hat H_0$, and solve for the steady state solution $\hat\rho^{(0)}\equiv \hat\rho_0^e$.
The elements $S^a$ are then simply found as the equilibrium expectation values $S^a = Av_0\langle \hat{S}^a({\bf r}_n) \rangle_{\hat \rho^{(0)}} = \mathrm{Tr} \{\hat{S}^a({\bf r}_n)\hat \rho^{(0)} \}$. Next, we add to the set of Bloch equations the first-order effect of a perturbation $\hat H' = \boldsymbol\Lambda \cdot \hat{\bf S}({\bf r}_n)$, i.e., we add the terms $i[\hat\rho^{(0)},\hat H']_-$, and solve for the new steady state density matrix $\hat \rho^{(1)}$. The spin susceptibility $\chi^{ab}$ needed in Eq.\ (\ref{eq:dkdtfin}) is then found as the part of $\langle \hat{S}^a({\bf r}_n) \rangle_{\hat \rho^{(1)}} = \mathrm{Tr} \{\hat{S}^a({\bf r}_n)\hat \rho^{(1)} \}$ which is linear in $\Lambda^b$. The fluctuations $R^{ab}$ are found in a similar way, the terms to add to the Bloch equations then read $[\hat\rho^{(0)},\hat H']_+ - 2\mathrm{Tr}\{\hat H'\hat\rho^{(0)}\}\hat\rho^{(0)}$, where the last term removes the connected part of the electron spin correlators. We will illustrate the procedure below.

From this short outline, it is already clear that this approach has several
properties which are different from 
a traditional FGR calculation: (i) For an FGR calculation, there should be an obvious basis to calculate all occupation probabilities in. Here we can choose any basis which is most convenient for evaluating the steady state electronic density matrix. (ii) When calculating separate FGR transition rates, it should a priori be clear which quantization axis to choose for the nuclear spins. In our formalism, as soon as $\hat \rho^{(1)}$ is found for the two types of perturbations, all elements of $\chi$ and $R$ can be read off immediately, yielding the dynamics for the full vector $\langle\hat{\bf K}_n\rangle$, not just the polarization along a single axis. (iii) For FGR to work, all levels involved in the hyperfine transitions (including possible virtual states in higher order processes) should be well separated in energy, i.e., the splittings should be larger than all decay rates and other possible incoherent processes present. In our method all incoherent processes are accounted for in the set of Bloch equations, we do not have to compare them with the other energy scales.
(iv) In an FGR calculation, when focusing on the nuclear spin dynamics of a small subsystem being part of a larger system (e.g.\ a quantum dot coupled to multiple reservoirs), one usually includes only the energy splittings inside the subsystem in the actual perturbation theory. All other processes relevant for the electron spin dynamics (such as the coupling to the reservoirs) have to be accounted for `afterward' in an effective density of states, which is not always possible. In our approach, we treat the coherent and incoherent parts of the dynamics in the subsystem on equal footing. One advantage is that the perturbation theory in our approach does not break down when the energy splittings in the subsystem vanish, as long as the energies corresponding to the incoherent dynamics are large enough.

Let us mention here also a drawback of our method. Our main result (\ref{eq:dkdtfin}) describes only the dynamics of average nuclear fields $\langle \hat{\bf K} \rangle$. Temporal fluctuations and other stochastic properties of the fields are not accessible via this approach. In an FGR calculation, one can separate the rates for flipping nuclear spins up and down, and derive from this separation a Fokker-Planck equation describing the stochastics of the nuclear fields~\cite{PhysRevLett.103.046601}.

\section{Example applications}

We will now illustrate the above with a calculation of the nuclear spin pumping in two example systems. First, we will consider nuclear spin pumping in a trivial toy model concerning a single quantum dot coupled to two leads. We will show how a smart implementation of FGR can produce an identical result in this case. Then we will investigate a more complicated setup, involving a double quantum dot, which cannot be dealt with in an FGR approach. We will show however how our formalism can be applied
without any problem, 
yielding a general equation for the nuclear spin pumping in this setup.

\subsection{Single dot: Fermi's golden rule with modifications}

\begin{figure}[t]
\includegraphics[width=8cm]{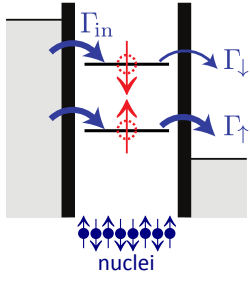}
\caption{Energy diagram for a trivial single quantum dot setup in which nuclear spin pumping is to be expected. The dot is coupled to two leads with different chemical potentials. One single-electron level lies within the bias window, which is split into two sublevels by the Zeeman energy (we assumed a negative $g$-factor). The tunnel rate from the left lead into the dot, $\Gamma_\text{in}$ is large and equal for both spin directions. The outward tunnel rates $\Gamma_{\ua,\da}$ are different for spin up and down.} \label{fig:fig1}
\end{figure}

We consider a single quantum dot connected to two leads, as illustrated in Fig.\ \ref{fig:fig1}. Due to the strong Coulomb repulsion, only one excess electron can occupy the dot. If we apply an external magnetic field ${\bf B}_\text{ext} = B_\text{ext}\hat z$, the energy levels for spin up and down in the dot are split by the electronic Zeeman energy. We take the tunnel rate into the dot $\Gamma_\text{in}$ to be very fast and equal for both spin directions. The outward tunnel rates $\Gamma_{\ua,\da}$ are different for the two spin directions (to provide a physical picture with this assumption: one could assume the right lead to be ferromagnetic). Since the electron occupying the dot will thus have on average a non-zero spin polarization along the $z$-direction, hyperfine interaction with the nuclear spins is expected to produce a non-zero nuclear spin polarization along this axis.

In typical experiments, the temperature ($\sim 100$~mK~$\sim 10~\mu$eV) is much smaller than the orbital level spacing in the dot ($\sim 1$~meV). In that case, the electron will in principle occupy the orbital ground state of the dot, which has an $s$-type character. As long as also the energy scale of the hyperfine interaction is much smaller than the orbital level spacing, we can project $\hat H_\text{hf}$ to the orbital ground state, and write
\begin{equation}
 \hat H_\text{hf}(t) = \sum_n A_n \hat S^a(t)\hat K^a_n,
\label{eq:hf2}
\end{equation}
where the hyperfine coupling coefficients are defined as $A_n=Av_0 |\psi_0({\bf r}_n)|^2$, i.e., in terms of the envelope wave function $\psi_0({\bf r})$ of the orbital ground state. The operator $\hat{{\bf S}}$ is now the electron spin operator, instead of spin density. A usual further simplification to make is to assume that $|\psi_0({\bf r}_n)|^2$ is approximately constant in the dot. This implies that $|\psi_0({\bf r}_n)|^2 \approx (Nv_0)^{-1}$, where $N\sim 10^4$--$10^6$ is the number of nuclear spins in the dot. In this case, all coupling coefficients simply reduce to $A_n = A/N$.

Let us now calculate the nuclear spin pumping rate along the $z$-direction, $d\langle \hat K_n^z \rangle /dt$, for this system. Since all nuclear spins are coupled to $\hat{\bf S}$ with the same coefficient, $\langle \hat K_n^z\rangle$ for the $n$-th nucleus equals the ensemble averaged nuclear spin $\langle \hat K^z\rangle$. We only take into account a finite polarization in the $z$-direction, so we use the equation
\begin{equation}
\frac{d\langle \hat K^z \rangle}{dt} = \frac{1}{3} P^z - \frac{1}{2}Q^{zz} \langle \hat K^z \rangle,
\label{eq:dkdtqd}
\end{equation}
with the functions $P^z = A_n^2 K(K+1) (\chi^{xy}-\chi^{yx})$ and $Q^{zz} = A_n^2 (R^{xx}+ R^{yy})$. Note that $\chi$ and $R$ contain now correlators of the electron spin, not electron spin density.

A finite nuclear polarization in the $z$-direction behaves on the time scale of the electronic dynamics as a static contribution to the magnetic field experienced by the electrons. We incorporate this contribution into an effective field $B_\text{eff} = B_\text{ext} - A\langle \hat K^z \rangle$ (we assume a negative $g$-factor and a positive hyperfine coupling constant, as is the case in GaAs). The coherent time-evolution of the electronic $2\times 2$ density matrix of this simple system is then given by $\partial_t \hat\rho = i[B_\text{eff}\hat S^z,\hat \rho]_-$, where we express $B_\text{eff}$ in terms of a frequency. To this we add
the incoherent tunnel rates $\Gamma_\ua$ and $\Gamma_\da$, which, to a good approximation, include the presence of the leads and the left and right tunnel barriers into the equations of motion for the electron on the dot. Together, this yields
the set of equations
\begin{equation}\label{masterne}
\begin{split}
\partial_t \rho_{\ua} &= -\Gamma_\uparrow \rho_{\ua} + \tfrac{1}{2}(\Gamma_\da \rho_{\da}+\Gamma_\ua\rho_{\ua}) \\
\partial_t \rho_{\da} &= -\Gamma_\da \rho_{\da}+ \tfrac{1}{2}(\Gamma_\da \rho_{\da}+\Gamma_\ua\rho_{\ua}) \\
\partial_t \rho_{\ua\da} &= (iB_\text{eff}-\tfrac{1}{2}\Gamma) \rho_{\ua\da}\\
\partial_t \rho_{\da\ua} &= (-iB_\text{eff}-\tfrac{1}{2}\Gamma) \rho_{\da\ua},
\end{split}
\end{equation}
where $\Gamma\equiv\Gamma_\ua+\Gamma_\da$.
These equations can be solved for the stationary situation $\partial_t\hat\rho^{(0)} = 0$, yielding $\rho^{(0)}_{\ua(\da)} = \Gamma_{\da(\ua)}/\Gamma$ and $\rho^{(0)}_{\ua\da}=\rho^{(0)}_{\da\ua}=0$. We only need the response functions $\chi^{xy}$ and $\chi^{yx}$, so we add the terms $i[\hat\rho^{(0)},\Lambda^x \hat S^x+\Lambda^y \hat S^y]_-$ to Eq.\ (\ref{masterne}) and solve for the new stationary solution $\hat \rho^{(1)}$. The correlator $\chi^{xy}$ is then simply given by the term in $\text{Tr}\{\hat S^x\hat\rho^{(1)}\}$ linear in $\Lambda^y$, and in a similar way we find $\chi^{yx}$. The results are combined to
\begin{equation}
\chi^{xy}-\chi^{yx} = \frac{\Gamma_\da - \Gamma_\ua}{2B_\text{eff}^2 + \frac{1}{2}\Gamma^2}.
\label{eq:chi}
\end{equation}
Now we evaluate $R^{xx}$ and $R^{yy}$. To this end, we add the terms $[\hat\rho^{(0)},\Lambda^x \hat S^x+\Lambda^y \hat S^y]_+- 2\mathrm{Tr} \{ (\Lambda^x \hat S^x+\Lambda^y \hat S^y)\hat \rho^{(0)}\}\hat\rho^{(0)}$ to Eq.\ (\ref{masterne}), and then look for the linear responses in $\hat S^x$ and $\hat S^y$, yielding
\begin{equation}
R^{xx}+ R^{yy} = \frac{\Gamma}{2B_\text{eff}^2 + \frac{1}{2}\Gamma^2}.
\label{eq:r}
\end{equation}
Combining all together using Eq.\ (\ref{eq:dkdtqd}), we thus find
\begin{equation}
\frac{d\langle \hat K^z \rangle}{dt} = A_n^2\frac{\Gamma_\da (\frac{1}{2}-\langle \hat K^z \rangle) - \Gamma_\ua(\frac{1}{2} + \langle \hat K^z \rangle)}{4(B_\text{ext} - A\langle \hat K^z \rangle)^2 + \Gamma^2},
\label{eq:result1}
\end{equation}
where for simplicity we assumed $K=1/2$.

Let us compare the result (\ref{eq:result1}) with the pumping rate we get using an FGR approach.
We focus on transitions between the two states $\ket{\ua\Downarrow_n}$ and $\ket{\da\Uparrow_n}$, where the double arrow indicates the nuclear spin along the $z$-axis. Neglecting the nuclear Zeeman energy, the two states are separated by the energy $\hbar B_\text{eff}$, and both are broadened by their respective electronic decay rate $\Gamma_{\ua(\da)}$. Hyperfine induced transitions from $\ket{\ua\Downarrow_n}$ to $\ket{\da\Uparrow_n}$ contribute {\it positively} to $d\langle\hat K_n^z\rangle/dt$, and their rate is given by
\begin{equation}\label{eq:fermi}
\Gamma_+ = 2\pi |\braket{\da\Uparrow_n|\tfrac{1}{2}A_n\hat S^-\hat K_n^+|\ua\Downarrow_n}|^2 p_i \, \mathcal{D},
\end{equation}
the square of the relevant matrix element, multiplied by the chance $p_i$ of finding the system initially in $\ket{\ua\Downarrow_n}$ and the effective `density of states' for this transition $\mathcal{D}$.

For the occupation probability $p_i$, we take $p_i = (\tfrac{1}{2}-\langle \hat K^z \rangle) \times \rho^{(0)}_{\ua}$. In this simple case, it is possible to incorporate all incoherent dynamics (the two decay rates) into a smartly chosen $\mathcal{D}$: We take the sum of the level broadenings of the two electronic levels $\ket{\ua}$ and $\ket{\da}$. Assuming a Lorentzian shape for these broadenings, we thus use
\begin{equation}
\mathcal{D} = \frac{1}{\pi}\frac{\frac{1}{2}\Gamma}{B_\text{eff}^2+\tfrac{1}{4}\Gamma^2},
\label{eq:dos}
\end{equation}
so that we can write
\begin{equation}
\Gamma_+ = A^2_n\frac{\Gamma_\da(\frac{1}{2}-\langle \hat K^z \rangle)}{4B_\text{eff}^2+\Gamma^2}.
\label{eq:gamma+}
\end{equation}
We derive similarly an equation for $\Gamma_-$ and then combine the two to write for the net nuclear spin pumping rate
\begin{equation}\label{eq:result2}
\frac{d\langle \hat K^z \rangle}{dt} = A_n^2 \frac{\Gamma_\da (\frac{1}{2}-\langle \hat K^z \rangle) - \Gamma_\ua(\frac{1}{2} + \langle \hat K^z \rangle)}{4(B_\text{ext} - A\langle \hat K^z \rangle)^2 + \Gamma^2}.
\end{equation}
We see that this result coincides with Eq.\ (\ref{eq:result1}). The fact that the FGR approach in this case also works for small energy splittings, i.e., in the regime $B_\text{eff} \lesssim \Gamma$, is due to the fact that all incoherent effects in this model can be incorporated relatively simply into $\mathcal{D}$.

\subsection{Double dot: Beyond Fermi's golden rule}

\begin{figure}[t]
\includegraphics[width=8cm]{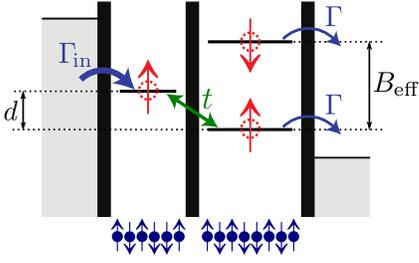}
\caption{Energy diagram of a more complicated setup. Two quantum dots are coupled to each other and to two leads with different chemical potentials. The two dots have different effective $g$-factors, so that the application of an external magnetic field yields different Zeeman energies in the dots. The double dot is tuned to the $(0,0)\to (1,0)\to (0,1)\to (0,0)$ transport regime. The outward tunneling rates $\Gamma$ are equal and much smaller than the inward rate $\Gamma_\text{in}$.}
\label{fig2}
\end{figure}
Let us now illustrate what can happen when the setup becomes more complicated. We consider a double quantum dot connected to two leads with different chemical potentials, as depicted in Fig.\ \ref{fig2}. The dots are tuned such that an electronic transport cycle involves the charge states $(0,0)\to (1,0) \to (0,1) \to (0,0)$, where $(n,m)$ denotes a state with $n(m)$ excess electrons on the left(right) dot. The electronic levels in the two dots are both split by a Zeeman energy in a spin up and spin down level. We assume that the splitting in the left dot is larger than in the right dot (e.g.\ due to a size-related difference in effective $g$-factors), and that the $(1,0)$ spin down level has a too high energy to play a role. The $(1,0)$ and $(0,1)$ spin up levels are detuned by an energy $d$ and coupled to each other with a tunnel coupling $t$. We again assume the outward tunneling rates to be much slower than the inward, $\Gamma \ll \Gamma_\text{in}$, so that the system effectively will never occupy the $(0,0)$ charge state.

We will focus on nuclear spin pumping in the right dot. A non-zero nuclear spin pumping rate is expected along the direction of the magnetic field since the average local electron spin polarization along this direction is finite. Denoting the level in the left dot by $\ket{L}$ and the two levels in the right dot by $\ket{\ua}$ and $\ket{\da}$, the Hamiltonian for the three-level system reads
\begin{equation}
\hat H_3 = d\ket{L}\bra{L} + B_\text{eff}\ket{\da}\bra{\da} + t\ket{L}\bra{\da}+t\ket{\da}\bra{L},
\label{eq:ham3}
\end{equation}
where the feedback of the nuclear spins on the electron spin dynamics again is incorporated into the effective magnetic field $B_\text{eff} = B_\text{ext} - A \langle\hat K^z\rangle$.
Using this Hamiltonian and adding the effect of the incoherent decay rate $\Gamma$, we can write down the equations of motion for $\hat \rho$, and solve for $\hat\rho^{(0)}$, yielding
\begin{equation}\begin{split}
\rho^{(0)}_L & = \frac{4d^2+4t^2+\Gamma^2}{4d^2+8t^2+\Gamma^2}, \quad \rho^{(0)}_\ua = \frac{4t^2}{4d^2+8t^2+\Gamma^2}, \\
\rho^{(0)}_{L\ua} & = (\rho^{(0)}_{\ua L})^* = \frac{4dt+2it\Gamma}{4d^2+8t^2+\Gamma^2},
\end{split}
\label{eq:rho03}
\end{equation}
and zero for the other five elements.
We then proceed as we did above, and add consecutively the perturbations $i[\hat\rho^{(0)},\hat H']_-$ and $[\hat\rho^{(0)},\hat H']_+ - 2\mathrm{Tr}\{\hat H'\hat\rho^{(0)}\}\hat\rho^{(0)}$ to the equations of motion. We solve in both cases for the new steady-state solution $\hat\rho^{(1)}$ and extract the linear responses needed. Combining all together, we find
\begin{widetext}
\begin{equation}
\frac{d\langle \hat K^z \rangle}{dt} = \frac{4 A_n^2 t^2 \Gamma  (B_\text{eff}^2+2 t^2+\Gamma^2)(\tfrac{1}{2}-\langle\hat K^z \rangle)}{(4 d^2+8 t^2+\Gamma^2) [4 (B_\text{eff}d-B_\text{eff}^2+t^2)^2+ (5 B_\text{eff}^2-8 B_\text{eff} d+4 d^2+4t^2) \Gamma ^2+\Gamma ^4]},
\label{eq:sol3}
\end{equation}
where $B_\text{eff} = B_\text{ext} - A \langle\hat K^z\rangle$. This equation gives the nuclear spin pumping in the right dot along the $z$-axis, without any restriction imposed on the relative magnitude of the parameters $B_\text{eff}$, $d$, $t$, and $\Gamma$.

Let us now try to evaluate the same pumping using an FGR approach. The most transparent electronic basis to use is $\{\ket{L},\ket{\ua},\ket{\da}\}$ since in this basis we have well-defined incoherent in- and outward tunneling rates. We see however that the stationary density matrix in this basis (\ref{eq:rho03}) has off-diagonal elements. In order to proceed, we have to make several assumptions. Let us consider the limit of $\Gamma \gg t$. In this case the off-diagonal elements of $\hat\rho^{(0)}$ can be neglected, and we can approximate $\hat\rho^{(0)} \approx \ket{L}\bra{L}$. We further assume that we have three well-separated levels, i.e., $B_\text{eff},d \gg \Gamma$. In this limit the nuclear spin pumping is resulting from a second order transition: (i) tunneling from the initial state $\ket{L\Uparrow_n}$ to the virtual state $\ket{\da\Uparrow_n}$, and then (ii) an electron-nuclear spin flip-flop from $\ket{\da\Uparrow_n}$ to the final state $\ket{\ua\Downarrow_n}$. We thus use a standard second order Fermi golden rule to write for the pumping rate
\begin{equation}
\frac{d\langle \hat K^z \rangle}{dt} = 2\pi \frac{|\braket{\da\Uparrow_n|\tfrac{1}{2}A_n\hat S^-\hat K_n^+|\ua\Downarrow_n}\braket{\ua\Downarrow_n|\hat H_T|L\Downarrow_n}|^2}
{(E_{\da\Uparrow_n}-E_{L\Uparrow_n})^2}
p_{L\Uparrow_n}\mathcal{D},
\label{eq:pump2o}
\end{equation}
\end{widetext}
where $\hat H_T = t\ket{L}\bra{\da}+t\ket{\da}\bra{L}$ is the tunneling part of the Hamiltonian $\hat H_3$. The occupation probability of the initial state is $p_{L\Uparrow_n} = \tfrac{1}{2} - \langle  \hat K^z \rangle$. The `density' $\mathcal{D}$ for this transition is mainly set by the level broadening of the final state, so $\mathcal{D} = \tfrac{2}{\pi}\Gamma / [4(B_\text{eff}-d)^2+\Gamma^2]$. This yields
\begin{equation}
\frac{d\langle \hat K^z \rangle}{dt} = \frac{A_n^2t^2\Gamma(\tfrac{1}{2}-\langle\hat K^z\rangle )}{d^2[4(B_\text{eff}-d)^2+\Gamma^2]},
\label{eq:sol4}
\end{equation}
which coincides with (\ref{eq:sol3}) in the limit $B_\text{eff},d \gg \Gamma \gg t$. (Note that $B_\text{eff}-d \gg \Gamma$ is not a necessary condition.)

There are other limits in which considering the right rates to calculate, combined with a good approximation for $\hat \rho^{(0)}$ and a smart choice for $\mathcal{D}$, can result in the right pumping rate. In general however, it is not trivial to understand how to incorporate all incoherent dynamics into $\mathcal{D}$. Even if one would simply diagonalize the stationary density matrix $\hat\rho^{(0)}$ and try to evaluate all hyperfine transitions in the new basis, one would have to transform the incoherent processes in correct effective densities of state. In the approach using the electron spin susceptibility and fluctuations, this is all done on the fly and its validity is not restricted to certain limits of the parameter space.

Let us emphasize here that the fact that a correctly adapted FGR approach did work in our single dot example but not in the double dot example, is only due to the complexity of the coherent and incoherent electron spin dynamics in the two cases. The success of the FGR approach is not inherently connected to the number of quantum dots in a setup: Also for more complicated single dot systems, the FGR approach might break down.

The two examples given above both consider localized electron spins. Our approach however, works in principle equally well for systems with delocalized electrons. In this case, as explained above, the susceptibility and fluctuations needed are correlators of electron spin {\it density} instead of simply electron spin. This however does not change the complexity of the formalism: One only has to find the right set of (Bloch) equations to describe the electron spin density.

\section{Conclusions}

To conclude, we presented a
new 
approach to evaluate the nuclear spin dynamics driven by hyperfine interaction with non-equilibrium electron spins. To describe the dynamics up to second order in the hyperfine coupling, it suffices to evaluate the susceptibility and fluctuations of the electron spin. This approach does not rely on a separation of electronic energy scales or the specific choice of electronic basis states, thereby overcoming practical problems which may arise
under certain circumstances
when using a more traditional formalism based on rate equations.

\section{Acknowledgments}

J.~D.~would like to thank P.~W.~Brouwer for helpful discussions. This work was supported by the Alexander von Humboldt Foundation and the `Stichting voor Fundamenteel Onderzoek der Materie'.

\section{Appendix: Derivation of Eq.~(\ref{eq:2ohf})}

Here we will show how to derive the time evolution equation (\ref{eq:2ohf}) using second order perturbation theory in the interaction picture.
The total Hamiltonian of the system $\hat H$ is split  into an unperturbed part $\hat H_0$ and a small perturbation $\hat H'$. In the interaction picture, all operators acquire a time-dependence governed by $\hat H_0$, so that for any operator $\hat A$ we have $\hat A_I(t) = e^{i\hat H_0 t}\hat Ae^{-i\hat H_0t}$ (we assumed for simplicity a time-independent $\hat H_0$). The remaining part of the total Hamiltonian, the perturbation $\hat H'$, governs the time-dependence of the wave function, $i\partial_t\ket{\psi_I} = \hat H_I'(t)\ket{\psi_I}$, where $\hat H_I'(t)$ is the Hamiltonian of the perturbation in the interaction picture.
The time-evolution operator for $\ket{\psi_I}$ thus reads
\begin{equation}
 \hat U_I(t,t') = \mathcal{T}\left[ \exp \left\{-i \int_{t'}^t\hat H_I'(\tau) d\tau\right\}\right],
\label{eq:timeevolpert}
\end{equation}
where $\mathcal{T}$ is the time-ordering operator. The relation between $\hat U_I$ and the `regular' time-evolution operator in the Schr\"{o}dinger picture $\hat U$ is straightforward to derive and reads $\hat U(t,t') = e^{-i\hat H_0t}\hat U_I(t,t')e^{i\hat H_0t'}$.

Let us now consider the evolution of the density matrix of a system described by the Hamiltonian $\hat H_0 +\hat H'$. We assume that at some time in the past $t_0 \to -\infty$ the system was still unperturbed and could be described by a stationary density matrix $\hat \rho_0$ which is determined solely by $\hat H_0$. The perturbation is then switched on adiabatically, formally done by replacing $\hat H' \to e^{\eta t}\hat H'$, where $\eta$ is an infinitesimally small positive real number. The density matrix at any later time $t$ can then be written as $\hat \rho(t) = \hat U(t,t_0)\hat \rho_0 \hat U(t_0,t)$, where $\hat U$ describes the time-evolution due to $\hat H_0 + e^{\eta t}\hat H'$. In terms of evolution in the interaction picture, we see that we can write
\begin{equation}
 \hat \rho(t) = e^{-i\hat H_0t}\hat U_I(t,t_0)\hat \rho_0 \hat U_I(t_0,t')e^{i\hat H_0t},
\end{equation}
where we used that $e^{i\hat H_0t_0}\hat \rho_0e^{-i\hat H_0t_0} = \hat\rho_0$ since $\hat\rho_0$ is per definition stationary under $\hat H_0$.

Introducing $\hat\rho_I (t) = e^{i\hat H_0 t}\hat \rho (t)e^{-i\hat H_0t}$, the `interaction' density matrix, we note that it obeys the simple relation
\begin{equation}
 \hat \rho_I(t) = \hat U_I(t,t_0)\hat \rho_0 \hat U_I(t_0,t'),
\label{eq:dridt}
\end{equation}
and that for the expectation value of any observable
\begin{equation}
 \langle A(t) \rangle = \mathrm{Tr} \{ \hat A \hat\rho(t) \} = \mathrm{Tr} \{ \hat A_I(t)\hat\rho_I (t) \}.
\end{equation}
From Eqs (\ref{eq:dridt}) and (\ref{eq:timeevolpert}) it follows immediately that we can write as time-evolution equation for $\hat\rho_I(t)$
\begin{equation}
 \frac{d\hat\rho_I}{dt} = -i [ \hat H'_I(t),\hat\rho_I(t)]_- .
\label{eq:timeeq}
\end{equation}
If one knows the stationary density matrix $\hat\rho_0$, one can iteratively use Eq.~(\ref{eq:timeeq}) to approximate
\begin{equation}
\begin{split}
  \frac{d\hat\rho_I}{dt} = & - i [ \hat H'_I(t),\hat\rho_0]_-\\
& - \int^t [ \hat H_I'(t),[ \hat H_I'(t'),\hat \rho_0 ]_- ]_- dt' + \dots
\end{split}
\label{eq:dridtexp}
\end{equation}

In our case, we would like to obtain a time-evolution equation for a {\it reduced} density matrix, i.e., separate the system into two subsystems, $\mathcal{S}^{(1)}$ and $\mathcal{S}^{(2)}$, and trace out the degrees of freedom of one of the subsystems. We assume that initially, at $t_0$, the density matrix can be separated, $\hat\rho_0 = \hat\rho_0^{(1)}\otimes\hat\rho_0^{(2)}$, where the superscripts label the subsystems. We then trace in Eq.~(\ref{eq:dridtexp}) over the degrees of freedom of $\mathcal{S}^{(2)}$, which yields up to second order in $\hat H'$ for the reduced density matrix
\begin{equation}
\begin{split}
  \frac{d\hat\rho^{(1)}_I}{dt} = & - i [ \mathrm{Tr}_2 \{ \hat H'_I(t) \hat\rho_0^{(2)}\},\hat\rho^{(1)}_0]_-\\
& -\!\! \int^t \!\!\mathrm{Tr}_2\big\{ [ \hat H_I'(t),[ \hat H_I'(t'),\hat \rho_0^{(1)}\!\otimes\hat \rho^{(2)}_0 ]_- ]_- \big\}dt'.
\end{split}
\label{eq:dridtexpred}
\end{equation}
In this expression, tracing out the degrees of freedom of $\mathcal{S}^{(2)}$ in general introduces correlation functions of those coordinates of $\mathcal{S}^{(2)}$ to which $\hat H'$ is coupled. If one separates the contributions of the connected and unconnected correlators thus appearing in this expansion, one can see that all not fully connected correlators are actually the higher order iteration terms for lower order perturbation terms, which can all be collected together by replacing $\hat\rho_0^{(1)} \to \hat\rho^{(1)}(t)$, similar to all terms in the iteration series (\ref{eq:dridtexp}) combining to (\ref{eq:timeeq}). In terms of only fully connected correlators of the coordinates of $\mathcal{S}^{(2)}$, we can thus write
\begin{equation}
\begin{split}
  \frac{d\hat\rho^{(1)}_I}{dt} = & - i [ \mathrm{Tr}_2 \{ \hat H'_I(t) \hat\rho_0^{(2)}\},\hat\rho^{(1)}_I(t)]_-\\
& -\!\! \int^t \!\!\overline{\mathrm{Tr}}_2\Big\{ [ \hat H_I'(t),[ \hat H_I'(t'),\hat \rho_I^{(1)}(t)\!\otimes\hat \rho^{(2)}_0 ]_- ]_- \Big\}dt',
\end{split}
\label{eq:dridtexpredt}
\end{equation}
which equals Eq.~(\ref{eq:2ohf}). The line over the trace symbol indicates that only fully connected contributions should be included here. Note that we dropped all `interaction' indices $I$ in the main text: all operators and density matrices are implicitly written in the interaction picture.


\begin{thebibliography}{10}

\bibitem{spinrev}
I.~Zuti\'{c}, J.~Fabian, and S.~{Das Sarma}, Rev. Mod. Phys. \textbf{76}, 323
  (2004).

\bibitem{wolfscience}
S.~A. Wolf, et al., Science \textbf{294},
  1488 (2001).

\bibitem{springerlink:10.1007}
G.~Prinz, in B.~Heinrich and J.~A.~C. Bland, eds., \emph{Ultrathin Magnetic
  Structures IV}, pp. 5--18, Springer Berlin Heidelberg (2005).

\bibitem{qdqb}
D.~Loss and D.~P. DiVincenzo, Phys. Rev. A \textbf{57}, 120 (1998).

\bibitem{ronaldrev}
R.~Hanson, et al., Rev. Mod. Phys. \textbf{79}, 1217 (2007).

\bibitem{frank:science}
F.~H.~L. Koppens, et al., Science \textbf{309}, 1346 (2005).

\bibitem{petta:science}
J.~R. Petta, et al., Science \textbf{309}, 2180
  (2005).

\bibitem{klg}
A.~V. Khaetskii, D.~Loss, and L.~Glazman, Phys. Rev. Lett. \textbf{88}, 186802
  (2002).

\bibitem{reillyt2}
D.~J. Reilly, et al., Science \textbf{321}, 817 (2008).

\bibitem{PhysRevLett.102.057601}
L.~Cywi\ifmmode~\acute{n}\else \'{n}\fi{}ski, W.~M. Witzel, and S.~Das~Sarma,
  Phys. Rev. Lett. \textbf{102}, 057601 (2009).

\bibitem{onosb}
K.~Ono and S.~Tarucha, Phys. Rev. Lett. \textbf{92}, 256803 (2004).

\bibitem{greilich:science}
A.~Greilich, et al., Science \textbf{317}, 1896 (2007).

\bibitem{PhysRevLett.97.146602}
J.~Danon and Yu.~V. Nazarov, Phys. Rev. Lett. \textbf{97}, 146602 (2006).

\bibitem{opticalorientation}
F.~Meier and B.~P. Zakharchenya, eds., \emph{Optical orientation}, Elsevier,
  Amsterdam (1984).

\bibitem{dyakonovbook}
M.~I. Dyakonov, ed., \emph{Spin physics in semiconductors}, Springer Berlin
  Heidelberg (2008).

\bibitem{tartakovskii:prl}
A.~I. Tartakovskii, et al., Phys. Rev. Lett. \textbf{98},
  026806 (2007).

\bibitem{Rudner}
M.~S. Rudner and L.~S. Levitov, Phys. Rev. Lett. \textbf{99}, 246602 (2007).

\bibitem{ivo:nature}
I.~T. Vink, et al., Nature Physics \textbf{5}, 764 (2009).

\bibitem{PhysRevLett.103.046601}
J.~Danon, et al., Phys. Rev. Lett. \textbf{103}, 046601 (2009).

\bibitem{pfund:prl}
A.~Pfund, I.~Shorubalko, K.~Ensslin, and R.~Leturcq, Phys. Rev. Lett.
  \textbf{99}, 036801 (2007).

\bibitem{PhysRevB.80.041301}
J.~Danon and Yu.~V. Nazarov, Phys. Rev. B \textbf{80}, 041301 (2009).


\bibitem{footnote2}
Note that we included here the unconnected correlators $\langle \hat K_n^d\rangle \langle\hat K_n^c \rangle$ from the third term in (\ref{eq:dkdt3}). The summation over $m$ includes the term $m=n$ in this paragraph.
 
\end{thebibliography}

\end{document}